\documentclass [12pt]{iopart}
\usepackage {iopams}
\usepackage{setstack}
\usepackage{amstext}
\usepackage{amsfonts}
\usepackage{graphicx}
\usepackage{subfigure}
\usepackage{array}
\usepackage{appendix}

\usepackage{mathrsfs}
\usepackage{color}
\usepackage{makecell}
\usepackage{adjustbox,lipsum}

\usepackage{algorithm}
\usepackage{algpseudocode}

\usepackage{soul}

\begin{document}

\title[]{$T$-depth-optimized Quantum Search with Quantum Data-access Machine}

\author{{Jung~Jun~Park}$^{1}$, {Kyunghyun~Baek}$^{2}$, {M.~S.~Kim}$^{3,4}$, {Hyunchul~Nha}$^{5}$, {Jaewan~Kim}$^{4}$, {Jeongho~Bang}$^{2}$}

\address{$^1$ LG Electronics, AI Lab, CTO Div, Seoul, 06772, Korea}
\address{$^2$ Electronics and Telecommunications Research Institute, Daejeon 34129, Korea}
\address{$^3$ QOLS, Blackett Laboratory, Imperial College London, London SW7 2AZ, United Kingdom}
\address{$^4$ School of Computational Sciences, Korea Institute for Advanced Study, Seoul 02455, Korea}
\address{$^5$ Department of Physics, Texas A\&M University at Qatar, POBox 23874, Doha, Qatar}

\vspace{10pt}

\begin{indented}
\item The first two authors (J.J.P. and K.B.) contributed equally to this study and can be regarded as the main authors. 
\item Correspondence and requests for materials should be addressed to J.K. and J.B.
\end{indented}

\ead{\mailto{jaewan@kias.re.kr} and \mailto{jbang@etri.re.kr}}

\begin{abstract}
Quantum search algorithms offer a remarkable advantage of quadratic reduction in query complexity using quantum superposition principle. However, how an actual architecture may access and handle the database in a quantum superposed state has been largely unexplored so far; the quantum state of data was simply assumed to be prepared and accessed by a black-box operation---so-called oracle, even though this process, if not appropriately designed, may adversely diminish the quantum query advantage. Here, we introduce an efficient quantum data-access process, dubbed as quantum data-access machine (QDAM), and present a general architecture for quantum search algorithm. We analyze the runtime of our algorithm in view of the fault-tolerant quantum computation (FTQC) consisting of logical qubits within an effective quantum error correction code. Specifically, we introduce a measure involving two computational complexities, i.e.  quantum query and $T$-depth complexities, which can be critical to assess performance since the logical non-Clifford gates, such as the $T$ (i.e., $\pi/8$ rotation) gate, are known to be costliest to implement in FTQC. Our analysis shows that for $N$ searching data, a QDAM model exhibiting a logarithmic, i.e., $O(\log{N})$, growth of the $T$-depth complexity can be constructed. Further analysis reveals that our QDAM-embedded quantum search requires $O(\sqrt{N} \times \log{N})$ runtime cost. Our study thus demonstrates that the quantum data search algorithm can truly speed up over classical approaches with the logarithmic $T$-depth QDAM as a key component.
\end{abstract}

\maketitle

\newtheorem{theorem}{Theorem}
\newtheorem{lemma}{Lemma}
\newtheorem{definition}{Definition}
\newtheorem{result}{Result}
\newtheorem{estimation}{Resource Estimation (RE)}

\newcommand{\bra}[1]{\left<#1\right|}
\newcommand{\ket}[1]{\left|#1\right>}
\newcommand{\abs}[1]{\left|#1\right|}
\newcommand{\norm}[1]{\left|\!\left| #1\right|\!\right|}
\newcommand{\expt}[1]{\left<#1\right>}
\newcommand{\braket}[2]{\left<{#1}|{#2}\right>}
\newcommand{\commt}[2]{\left[{#1},{#2}\right]}
\newcommand{\round}[1]{\ensuremath{\lfloor#1\rceil}}

\newcommand{\identity}{1\!\!1}

\section{Introduction}\label{sec:intro}

Quantum search is one of the extolled quantum algorithms providing a profound insight into the quantum superposition principle with its quadratic quantum query advantage~\cite{Grover1997,Zalka1999,Bao2016}. A conventional scenario of the quantum search is to iterate a unitary operation, often-called quantum amplitude amplification (QAA) operation, to an initial state of the quantum-superposed data. The QAA operation contains a black-box known as the quantum oracle, which ``shifts'' the quantum phase of the state of the target data. The quantum amplitude of the target data is then increased while those of the other data decay as the QAA operation is iterated. In this sense, the algorithm is sometimes called amplitude amplification as a class name~\cite{Brassard2002,Ambainis2004}. In such a scheme, the $O(\sqrt{N})$ oracle calls (or quantum queries) are sufficient to find the target data among the unsorted $N$, whereas $O(N)$ is required in a classical search. Hence, quadratic improvement due to quantum superposition is achieved in the query complexity~\cite{Zalka1999}. 

However, it is unclear how the dataset to search is brought into a superposition state and how the phase-shift operation is performed by an oracle on a specific target state. That is, the {\em quantum data-access} process has not been properly considered. Here, quantum data-access means that the system selects and calls the data from a chosen range of the database or memory in the form of quantum superposition\footnote{In a broad context, ``data-access'' refers to the authority to access and control the data within a database or separate storage. Thus, such a description of the quantum data-access may be limited. However, since the concrete implementations of the (virtual) quantum gadget, such as database or a separate storage in a quantum computer, are yet to be known, it is not possible to directly extend a common definition of data-access to quantum data-access.}. Therefore, designing or using an efficient quantum data-access process is crucial as an ill-designed data access may offset the achieved quadratic quantum advantage~\cite{Viamontes2005}. This data-perspective issue has only recently been discussed in depth, for example, with some applications to the quantum machine learning~\cite{Aaronson2015,Tang2021}. Similar problems have been noted, and several conceptual studies have been conducted using a simplified framework~\cite{Dhawan2011,Broda2016,Kain2021}. Thus, developing an advanced model of the quantum data-access process, which we dub the quantum data-access machine (QDAM), is critically required. After all, it is important to analyze whether quadratic quantum speedup can consistently be realized in terms of the actual algorithm runtime, which necessarily hinges on the efficiency of the QDAM.

This work considers a general architecture of the search algorithm in which the QDAM can efficiently access to the (user-recognizable) searching data and connect them to a superposition state. The runtime of our algorithm is analyzed in the framework of the fault-tolerant quantum computation (FTQC) by assuming that an efficient quantum error correction (QEC) code is embedded in the logical qubits. In a FTQC model, the logical gate operations consist of two gate groups: Clifford and non-Clifford gates. In a popular scenario of FTQC, the Clifford gates, e.g., Hadamard, $\pi/4$-rotation, and controlled-NOT (CNOT), can be executed ensuring that the errors do not spread between the qubits; a transversal (or block-wise) run is set. On the other hand, in such a setting, an effective transversal implementation is impractical for the logical non-Clifford gates, typically $T$ gate, or equivalently $\pi/8$-rotation, and they are executed by preparing the so-called fault-tolerant magic state\footnote{While such a scenario is common, it is not the sole option available for FTQC. A transversal implementation of $T$ gate can be possible in theory, even though there is no code that utilizes both transversal and universal for every gates~\cite{Gottesman1997,Eastin2009}.}. However, the preparation of the magic state also causes errors, and  we should filter the noisy magic states into fewer high quality ones for an FTQC. This procedure, known as the magic state distillation, requires much more computation time and resources compared to running other logical Clifford gates~\cite{Bravyi2005,Horsman2012}. Thus, in FTQC, both reducing $T$-depth (a number of layers of the $T$ gates) and $T$-count (a total number of $T$ gates) are important for saving computational resources, as the implementation of $T$ gates with the magic state distillation is probabilistic~\cite{Fowler2012qec,Amy2014,Paler2017}. However, from the perspective of minimizing the {\em runtime} resources, $T$-depth holds significant importance\footnote{For instance, for the parallelized $T$ gates, the runtime mainly depends on one $T$ gate requiring the largest number of additional (e.g., $S$ or $Z$) corrections. On the other hand, in the case of sequentially arranged $T$'s, the corrections of all $T$ gates collectively affect the overall runtime.}. In this context,  we introduce a reasonable measure of runtime cost using both of  the quantum query and $T$-depth complexities. Through our analysis, we construct an efficient QDAM that can exhibit a logarithmic, i.e., $O(\log{N})$, $T$-depth, while (more than) $O(N)$ $T$-depth would otherwise be inevitable. Further analysis shows that it is possible to design an $O(\sqrt{N} \times \log{N})$ runtime quantum search with our logarithmic $T$-depth QDAM; otherwise, $O(\sqrt{N} \times N)$ is the limit, which would make a quantum search algorithm useless. Therefore, we conclude that the quadratic reduction in the query complexity, which has been considered as the quantum advantage thus far, can indeed lead to the runtime speedup in an actual architecture.

\section{Problem setting}\label{sec:pb_set}

\begin{figure}
\centering
\includegraphics[width=0.60\textwidth]{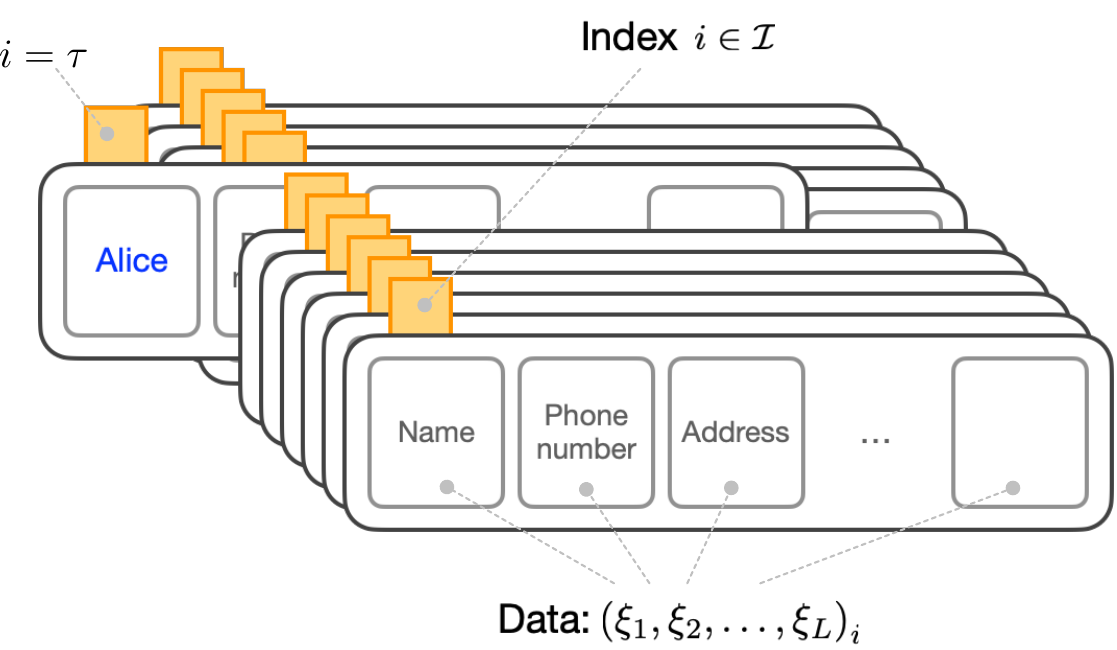}
\caption{Schematic of the grouped data indexed by $i \in {\cal I}$. Note that the indices $i$ are uncomputable. The data are unsorted.}
\label{fig:structured_data} 
\end{figure}

First, we consider a set (or table), say $\Xi_D$, of the unsorted data as
\begin{eqnarray}
\Xi_D = \left\{ \left(\xi_1, \xi_2, \ldots, \xi_L \right)_i \right\},
\end{eqnarray}
where the set of $L$-grouped data is indexed by $i \in {\cal I}$. Here, ${\cal I}$ is a finite index set referring to each subject to be searched, i.e. $i=1,\cdots,N$. Note that the indices $i$ are uncomputable quantities and the data are usually structured by certain known boundaries~\cite{Blakeley1996}. As an example, we consider personal data, consisting of a name (denoted by $\xi_1$), phone number (denoted by $\xi_2$), and address (denoted by $\xi_3$), {\it etc}. for each person $i$ (see Fig.~\ref{fig:structured_data}). We note that the data $\xi_k$ should be user-recognizable, or equivalently, deterministic. One can then generate a natural form of the command, {\em Find Alice's phone number.} This command can be translated into the problem of identifying the {\em unknown} index $\tau \in {\cal I}$ of the given $(\xi_1)_\tau=\text{`Alice.'}$ After identifying $\tau$, the task is accomplished by returning $(\xi_2)_\tau$, i.e., Alice's phone number. 

The search problem can thus be defined as follows:
\begin{definition}
Given $\Xi_D$ and a specific piece of data $(\xi_\alpha)_\tau$, the task is to identify the unknown index $\tau$.
\label{def:d_search}
\end{definition}
For simplicity, we assume that $(\xi_\alpha)_i \neq (\xi_\alpha)_{i'}$ for $i \neq i'$. Such a problem formulation follows a common scenario of the data-processing tasks~\cite{Pramanik1986,Mishra1992}. 

As the data $(\xi_k)_i$ are user-recognizable, we can define a ``deterministic'' state of the data register as~\cite{Park2019}
\begin{eqnarray}
\ket{\Xi_D} = \prod_{i \in {\cal I}} \ket{\left( \xi_1, \xi_2, \ldots, \xi_L \right)_i},
\end{eqnarray}
where $\ket{\left( \xi_1, \xi_2, \ldots, \xi_L \right)_i} = \ket{(\xi_1)_i}\ket{(\xi_2)_i}\cdots\ket{(\xi_L)_i}$ with the cardinality $\abs{\cal I} = N$. Here, $\ket{\Xi_D}$ is not a superposed state, representing a deterministic nature. The ``measured'' states $\ket{\left( \xi_1, \xi_2, \ldots, \xi_L \right)}$ are indexed by the symbols $i \in {\cal I}$, but are unsorted. Note that the use of such a quantum register, which is at least as large as the number of searching data (i.e., $O(\abs{\cal I}) = O(N)$) is essential, regardless of the efficiency of the QDAM.

We then define the three systems of the (logical) qubits with different purposes as
\begin{eqnarray}
\ket{\mathbf{q}}&=&\ket{q_1 q_2 \cdots q_n}~(\text{binary-encoded index qubits}), \nonumber \\
\ket{\mathbf{u}}&=&\ket{u_1 u_2 \cdots u_{2^n}}~(\text{one-hot-encoded index qubits}), \nonumber \\
\ket{\mathbf{d}}&=&\ket{d_1 d_2 \cdots d_m}~(\text{data qubits}), 
\label{eq:qb_registers}
\end{eqnarray}
where $\ket{q_j}, \ket{u_j}, \ket{d_j} \in \left\{ \ket{0}, \ket{1} \right\}$ and we let $2^n \le \abs{\cal I} = N$. The system of the binary-encoded index consists of $n$ qubits. These states, denoted as $\ket{\mathbf{q}}$, are represented by a sequence of binary numbers $\mathbf{q} = q_1 q_2 \cdots q_n$, where $q_j \in \{0, 1\}$. We also use a $2^n$-qubit one-hot-encoded index system. In this case, the states, denoted as $\ket{\mathbf{u}}$, are represented by the unary numbers $\mathbf{u}=u_1 u_2 \cdots u_{2^n}$, where $u_j = \delta_{j \kappa}$ for a specific $\kappa$. The data system brings (or copies) the data in $\ket{\Xi_D}$ into the $m$-qubit states, denoted as $\ket{\mathbf{d}}$. This process may allow only the specific part of the database corresponding to the given data (i.e. $|\xi_\alpha\rangle$) or the entire data (i.e. $|\xi_1,\ldots,\xi_\alpha,\ldots,\xi_L\rangle$) to be handled. In the former case, we can economize the qubit size (i.e., $m$, of the data system); however, we must call $\ket{\Xi_D}$ once more after completing the algorithm to access the target data (e.g., $(\xi_2)_\tau = \text{Alice's `phone number'}$) with the identified $\tau$. The latter returns all data connected to $\tau$, despite consuming larger qubits. These two encodings of the data state do not make any difference in terms of the overall computation costs if $L$ is not significantly large, i.e., $L \ll N$. We thus consider the former throughout this work without loss of generality.

\section{QDAM-embedded quantum search}\label{sec:alg}

\begin{figure}
\centering
\includegraphics[width=0.70\textwidth]{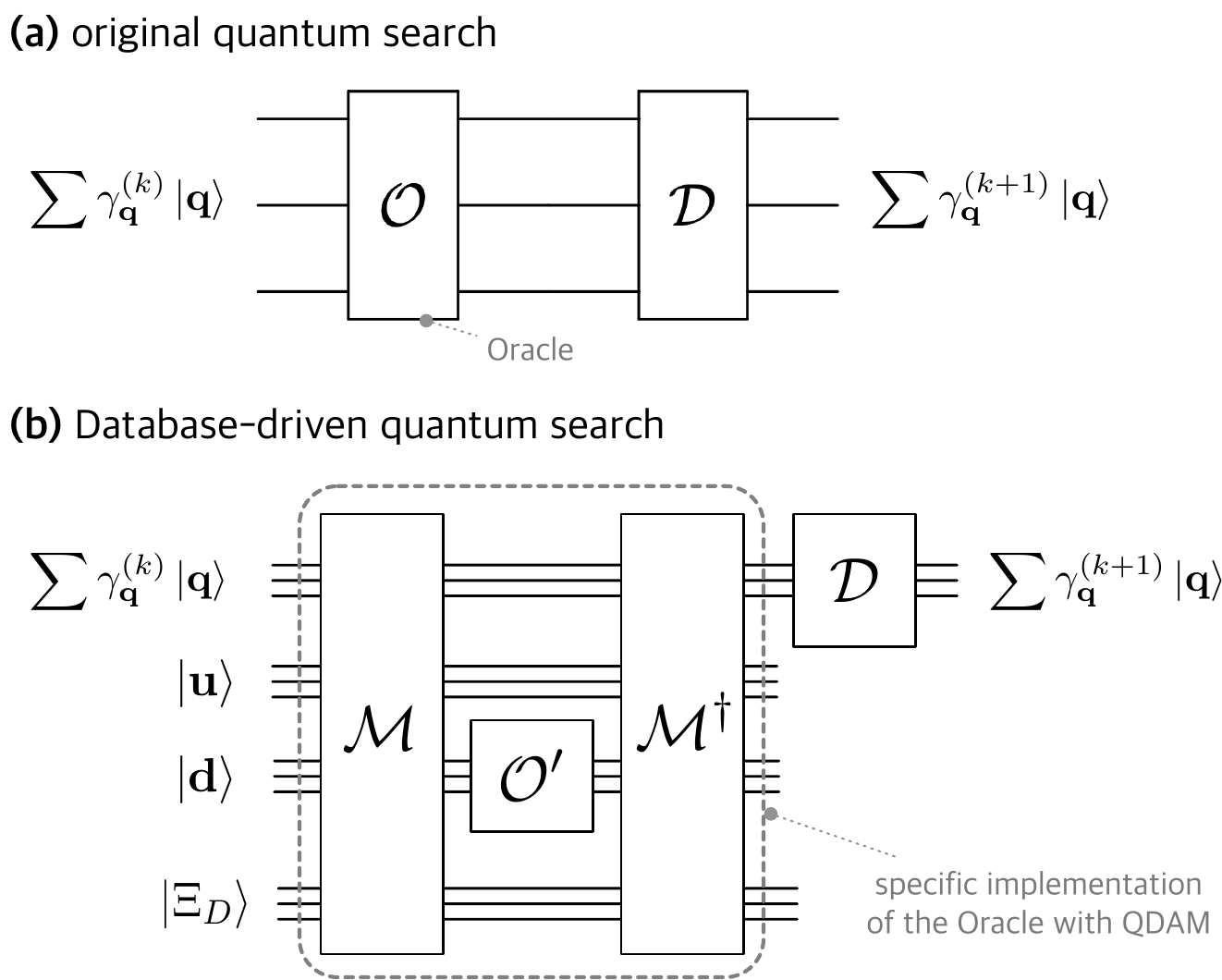}
\caption{Single-iteration QAA of the (a) original quantum search and (b) redesigned quantum search with QDAM.}
\label{fig:D_data_search} 
\end{figure}

We now start designing the quantum search algorithm to complete the task of Def.~\ref{def:d_search}. The algorithm consists of several subroutines: QDAM (denoted by ${\cal M}$), target reflection (denoted by ${\cal O}'$), the inverse of the $\text{QDAM}$ (denoted by ${\cal M}^\dagger$), and initial-state reflection (denoted by ${\cal D}$). In the original quantum search, the quantum oracle is conceptually considered as ${\cal O}'$ without a detailed description~\cite{Grover1997}. In our analysis, ${\cal M}$ (${\cal M}^\dagger$) is responsible for the connection (disconnection) between the encoded indices and the corresponding data. The structural differences between the original algorithm and our extended one with the QDAM are depicted in Fig.~\ref{fig:D_data_search}.

Our quantum search algorithm goes as follows:
\begin{algorithm}[H]
\caption{QDAM-embedded quantum search}\label{DQS}
\begin{algorithmic}[1]
\Require{$\ket{\Xi_D}$, $\ket{\mathbf{q}}$, $\ket{\mathbf{u}}$, $\ket{\mathbf{d}}$, ${\cal M}$ (QDAM), ${\cal O}'$, and ${\cal D}$}
\State Task: Given $(\xi_\alpha)_\tau$, find $(\xi_\beta)_\tau$~\Comment{$\xi_\beta$ and $\tau$ are unknown}
    \State Set ${\cal S} = \{ \mathbf{q} \}$~\Comment{$\abs{\cal S} \le \abs{\cal I}$}
    \State Initialize $\ket{\Psi} \gets \frac{1}{\sqrt{\abs{\cal S}}}\sum_{\mathbf{q} \in {\cal S}}\ket{\mathbf{q}}, \ket{\text{null}(\mathbf{u})}, \ket{\text{null}(\mathbf{d})}$
    \For{$k=1$ To $K=O(\sqrt{\abs{\cal S}})$}
        \State $\ket{\Psi} \gets {\cal M}^\dagger {\cal O}' {\cal M}\ket{\Psi}\ket{\Xi_D}$
        \State $\ket{\Psi} \gets {\cal D}\ket{\Psi}$
    \EndFor
    \State Identify $\ket{\tilde{\mathbf{q}}}$ by measuring the binary qubit system
    \State Obtain a candidate index $\tilde{\tau} \leftrightarrow \tilde{\mathbf{q}}$
    \If{$\ket{(\xi_\alpha)_{\tilde{\tau}}} = \ket{(\xi_\alpha)_{\tau}}$}
    \State Confirm $\tilde{\tau} = \tau$ \Comment{$\tau$ is achieved}
    \State \Return $\ket{(\xi_\beta)_{\tau}}$ by ${\cal M}$ \Comment{Solution}
    \Else
    \State \Return ``Algorithm failure''
    \EndIf
\end{algorithmic}
\end{algorithm}

To initiate, we select a set ${\cal S} = \{ \mathbf{q} \}$, each of whose element $\mathbf{q}=q_1 q_2 \ldots q_n$ is to be mapped to an index $i \in {\cal I}$. Generally, $\abs{\cal S}=2^n \le \abs{\cal I}=N$. However, it can be assumed to be $N=2^n$ in the extreme case. Subsequently, we prepare an index-superposed state, such that
\begin{eqnarray}
\ket{\psi_0} = \frac{1}{\sqrt{\abs{\cal S}}} \sum_{\mathbf{q} \in {\cal S}} \ket{\mathbf{q}}.
\label{eq:psi_0}
\end{eqnarray}

(i) {\em Initialization.}--- We start by initializing the states of the entire system as
\begin{eqnarray}
\ket{\Psi} = \ket{\psi_0} \ket{\text{null}(\mathbf{u})} \ket{\text{null}(\mathbf{d})},
\label{eq:Psi_i}
\end{eqnarray}
where $\ket{\text{null}(\mathbf{u})}$ and $\ket{\text{null}(\mathbf{d})}$ correspond to null states of the one-hot index and data qubits, respectively. At this initializing stage, $\ket{\psi_0}$, $\ket{\text{null}(\mathbf{u})}$, and $\ket{\text{null}(\mathbf{d})}$ are not correlated with each other.

(ii) {\em Index-data connection.}---We then use the QDAM, denoted by ${\cal M}$, to bring the $i$-indexed data $\ket{(\xi_\alpha)_i}$ into the data qubits by mapping $\{ \mathbf{q} \} = {\cal S}$ to $\{ i \} \subseteq {\cal I}$. Specifically, we generate a highly-entangled state by applying ${\cal M}$ on $\ket{\Psi}$ and $\ket{\Xi_D}$, as follows:\footnote{Here, $\ket{\psi_0}$ is first correlated with the unary qubits, and subsequently, $\ket{(\xi_\alpha)_i}$ are written into the data qubits through the unary qubits.}
\begin{eqnarray}
\left( \frac{1}{\sqrt{\abs{\cal S}}} \sum_{\mathbf{q} \in {\cal S}} \ket{\mathbf{q}, \mathbf{u}_\mathbf{q}, \mathbf{d}_{\mathbf{q} \leftrightarrow i}}\right),
\label{eq:Psi_1}
\end{eqnarray}
where $\mathbf{u}_\mathbf{q}$ denotes the one-hot representation of $\mathbf{q}$. When $n=2$, for instance, $\mathbf{u}_\mathbf{q}$ corresponding to $\mathbf{q}=00$, $01$, $10$, and $11$ are given as $1000$, $0100$, $0010$, and $0001$, respectively. Furthermore, $\mathbf{d}_{\mathbf{q} \leftrightarrow i}$ represents that the data $(\xi_\alpha)_i$ is linked to $\mathbf{q}$. Here, the one-hot index qubits are used to optimize the $T$-depth of the circuit of ${\cal M}$. The detailed design of the ${\cal M}$ and $T$-depth optimization will be provided later. 

(iii) {\em Target reflection.}---The next step marks the phase on the state $\ket{\boldsymbol\tau}\ket{\mathbf{u}_{\boldsymbol\tau}}\ket{\mathbf{d}_{\boldsymbol\tau}}$, where $\boldsymbol\tau \in {\cal S}$ corresponds to the target index $\tau \in {\cal I}$. To do this, we apply operation ${\cal O'}$, defined as
\begin{eqnarray}
\hat{\identity} - \left( 1 - e^{-i\phi} \right) \ket{\mathbf{d}_{\boldsymbol\tau}}\bra{\mathbf{d}_{\boldsymbol\tau}},
\label{eq:p-mark_op}
\end{eqnarray}
which performs a reflection about $\ket{\mathbf{d}_{\boldsymbol\tau}}$ with the phase $\phi$. This operation is often called Hilbert-space Householder reflection (HHR)~\cite{Ivanov2006}. In this work, $\phi$ is set to be $\pi$. Note that $\mathbf{d}_{\boldsymbol\tau}$ is assumed to be known as $(\xi_\alpha)_\tau$. After that, we have
\begin{eqnarray}
\frac{1}{\sqrt{\abs{\cal S}}} \left(  \sum_{\mathbf{q} \neq \boldsymbol\tau} \ket{\mathbf{q}, \mathbf{u}_\mathbf{q}, \mathbf{d}_\mathbf{q}} - \ket{\boldsymbol\tau, \mathbf{u}_{\boldsymbol\tau}, \mathbf{d}_{\boldsymbol\tau}}\right) \ket{\Xi_D}.
\label{eq:Psi_2}
\end{eqnarray}
Here, the phase marking can be accomplished without knowing $\boldsymbol\tau$, whereas it is unclear in the original algorithm how the black-box oracle marks the phase on the unknown target.

(iv) {\em Index-data disconnection.}---The three-qubit systems should be separated for the amplitude of $\ket{\boldsymbol\tau, \mathbf{u}_{\boldsymbol\tau}, \mathbf{d}_{\boldsymbol\tau}}$ to increase efficiently. We note that the binary index is advantageous for the main computation. Therefore, we apply the inverse QDAM, ${\cal M}^\dagger$. Then, the state in Eq.~(\ref{eq:Psi_2}) is decoupled as
\begin{eqnarray}
\frac{1}{\sqrt{\abs{\cal S}}} \left(  \sum_{\mathbf{q} \neq \boldsymbol\tau} \ket{\mathbf{q}} - \ket{\boldsymbol\tau}\right)\ket{\text{null}(\mathbf{u})}\ket{\text{null}(\mathbf{d})},
\label{eq:Psi_3}
\end{eqnarray}
which ensures that the remaining process is performed only on the binary-encoded index qubits. 

(v) {\em Initial-state reflection.}---The amplitude of the target index should then be increased, which is possible by reflecting the binary index state (tracing over the other states) in Eq.~(\ref{eq:Psi_3}) about the initial state $\ket{\psi_0}$. This task, denoted by ${\cal D}$, is also defined as HHR, such that
\begin{eqnarray}
\hat{\identity} - \left( 1 - e^{-i\varphi} \right) \ket{\psi_0}\bra{\psi_0}.
\label{eq:initial_HHR}
\end{eqnarray}
Here, a critical condition called the phase matching (i.e., $\phi = \varphi$) should be satisfied between ${\cal O}'$ and ${\cal D}$~\cite{Long1999,Li2002}. Throughout the work, we set $\varphi = \phi = \pi$. After that, the state of the binary index is
\begin{eqnarray}
\ket{\psi_1} = \sum_{\mathbf{q} \in {\cal S}} \gamma_\mathbf{q}^{(1)} \ket{\mathbf{q}},
\label{eq:psi_1}
\end{eqnarray}
where the following is satisfied:
\begin{eqnarray}
\abs{\gamma_{\boldsymbol\tau}^{(1)}} > \frac{1}{\sqrt{\abs{\cal S}}} > \abs{\gamma_{\mathbf{q \neq \boldsymbol\tau}}^{(1)}}.
\end{eqnarray}

The steps from (i)-(v) correspond to a single iteration of the quantum search. The algorithm is continued by going back to the step (i) using the achieved Eq.~(\ref{eq:psi_1}) and processing steps (i)-(v) again. We repeat (i)-(v) and halt the algorithm after $K$ repetitions. For each $k$th repetition, we may designate the binary index state (i.e., a generalized form of Eq.~(\ref{eq:psi_1})) as
\begin{eqnarray}
\ket{\psi_k} = \sum_{\mathbf{q} \in {\cal S}} \gamma_\mathbf{q}^{(k)} \ket{\mathbf{q}}.
\end{eqnarray}
If the halting number $K$ is set to be of the order $O(\sqrt{\abs{\cal S}})$, we can meet the following condition:
\begin{eqnarray}
\abs{\gamma_{\boldsymbol\tau}^{(0)}}^2 = \frac{1}{\abs{\cal S}} < \abs{\gamma_{\boldsymbol\tau}^{(1)}}^2 < \cdots < \abs{\gamma_{\boldsymbol\tau}^{(K)}}^2 \simeq 1.
\end{eqnarray}
After the completion of the algorithm, we obtain a candidate $\tilde{\mathbf{q}}$ ($\leftrightarrow \tilde{\tau}$), which is believed to be $\boldsymbol\tau$ ($\leftrightarrow \tau$). We then call ${\cal M}$ on the states $\ket{\tilde{\mathbf{q}}}\ket{\text{null}(\mathbf{u})}\ket{\text{null}(\mathbf{d})}\ket{\Xi_D}$ and measure the data state $\ket{\mathbf{d}_{\tilde{\mathbf{q}}}}$ to identify $\ket{(\xi_\alpha)_{\tilde{\tau}}}$. If the measured state is equal to $\ket{(\xi_\alpha)_\tau}$ (initially given), we confirm that $\ket{\tilde{\mathbf{q}}} = \ket{\boldsymbol\tau}$ and achieve $\tau$. Finally, the algorithm returns $(\xi_\beta)_\tau$ using ${\cal M}$ once again. In contrast, if the measured state is not equal to $\ket{(\xi_\alpha)_\tau}$, ``algorithm failure'' is returned. In this case, we have to run the algorithm again, assuming that $\tilde{\mathbf{q}} \neq \boldsymbol\tau$. If the failure continues, we try with the newly chosen ${\cal S}' \neq {\cal S}$, because such a situation could occur due to $\boldsymbol\tau \not\in {\cal S}$. 

\section{Analysis: runtime optimization}\label{sec:anal}

\subsection{Runtime cost for quantum search} 

Query complexity has been a useful measure with which the performance of quantum search algorithms is analyzed~\cite{Zalka1999}. The query complexity, say $Q$, is evaluated as
\begin{eqnarray}
Q = O\left(\frac{K}{1-\epsilon}\right),
\end{eqnarray}
where $\epsilon$ denotes the probability that the algorithm fails. Letting $\abs{\cal S} = N$ (as the extreme case), we conceive that $\epsilon \ll 1$ for large $N$. Then, we may obtain
\begin{eqnarray}
Q=O(K)=O(\sqrt{\abs{\cal S}}) = O(\sqrt{N}),
\label{eq:q_query_adv}
\end{eqnarray}
which ensures the quadratic quantum query advantage over the classical search requiring $O(N)$.

While the analysis of the quantum query complexity has been useful~\cite{Ambainis2004,Aaronson2021}, it is impractical to confirm the supremacy because the actual algorithm runtime is not on par with the query complexity. This is mainly because the (runtime) resources in the quantum data-access is not considered in the analysis. We thus need a more practical measure to assess the overall runtime of the computation, unlike the original version of the quantum search that incorporates only a black-box oracle. Here, we introduce a measure of the algorithm runtime taking FTQC into consideration,
\begin{eqnarray}
\text{$T$-$Cost$} = Q \times T_d,
\label{eq:cost}
\end{eqnarray}
where $T_d$ denotes the $T$-depth for a single run of the QAA, i.e. from (ii) to (v). The cost is analyzed in FTQC context as the QDAM is highly vulnerable to propagating errors with a large number of correlation gates like Toffoli gates~\cite{Sleator1995}. In FTQC,  Eq.~(\ref{eq:cost}) can be a reasonable measure to estimate the overall algorithm runtime because the non-Clifford gates, i.e., $\hat{T}$ and $\hat{T}^\dagger$, require significantly more resources and time to correct errors, as mentioned in Sec.~\ref{sec:intro}~(See Refs.~\cite{Fowler2012td,Amy2014,Paler2017,Niemann2019} for more details about the $T$-depth complexity).

In the situation where Eq.~(\ref{eq:q_query_adv}) is secured, $T_d$ should be optimized, which is of great importance: If $T_d$ is larger than $O(\sqrt{N})$, the $T$-$Cost$ becomes larger than $O(N)$, which is equivalent to those that can be achieved by a classical computer. Specifically, it is possible for Eq.~(\ref{eq:q_query_adv}) to not lead to the actual quantum speedup.

\subsection{Optimization and analysis of $T$-depth} 

Analysis of $T_d$ can be straightforwardly done by summing the $T$-depth of the subroutine machinery---${\cal M}$ (QDAM), ${\cal O}'$, and ${\cal D}$---that is,
\begin{eqnarray}
T_d = 2T_{d, {\cal M}} + T_{d, {\cal O}'} + T_{d, {\cal D}},
\label{eq:T_d}
\end{eqnarray}
where $T_{d, {\cal M}}$, $T_{d, {\cal O}'}$, and $T_{d, {\cal D}}$ denote the $T$-depth of ${\cal M}$, ${\cal O}'$, and ${\cal D}$, respectively. The factor $2$ of the $T_{d, {\cal M}}$ term arises because the $T$-depth of ${\cal M}^\dagger$ is equal to $T_{d, {\cal M}}$. From now on, we optimize $T_{d, {\cal M}}$, $T_{d, {\cal O}'}$, and $T_{d, {\cal D}}$.

\begin{figure}
\centering
\includegraphics[width=0.65\textwidth]{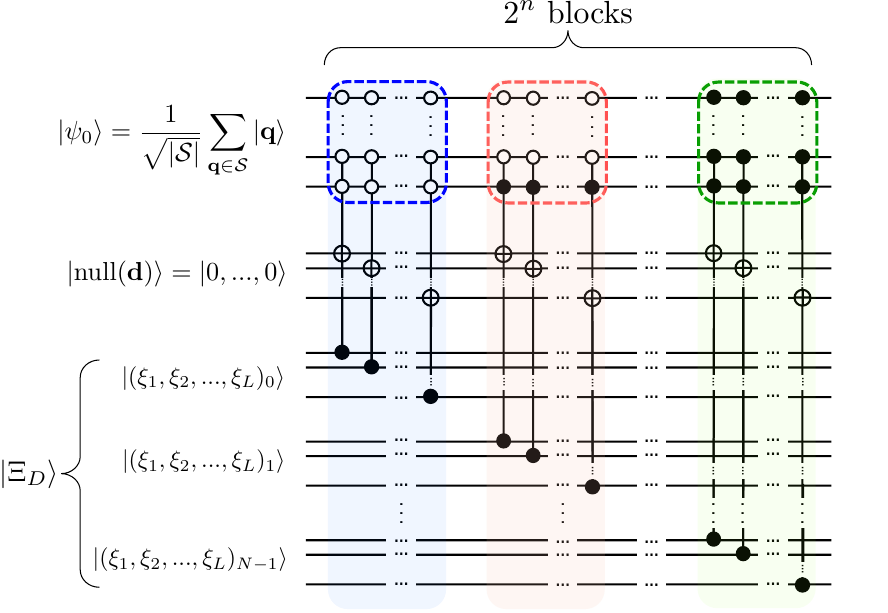}
\caption{A naive approach for the QDAM without $T$-depth optimization. In this approach, $m$ number of $(n+1)$-control-qubit Toffoli gates are required to encode each group of data, and a total number $|\mathcal S|$ of such gate operations are sequentially executed. Namely, $m\times |\mathcal S|$ number of $(n+1)$-Toffoli gates are needed.}
\label{fig:naive_QDAM} 
\end{figure}

{\em T-depth of ${\cal M}$.}---We first consider the $T$-depth of the QDAM, denoted as $T_{d, {\cal M}}$. QDAM here correlates the indices and the data in quantum superposition. A naive, simple, approach to achieve this task would be to apply $(n+1)$-control-qubit Toffoli gates---a generalization of the Toffoli gate with $n+1$ control qubits, where $n$ and $1$ refer to the index qubits and the data in $\Xi_D$, respectively (Fig.~\ref{fig:naive_QDAM}). However, this case requires a total number of $m \times 2^n$ execution of $(n+1)$-control-qubit Toffoli gates as the data encoding for each $i=1,\cdots,N$ is done sequentially all using the common control qubits  (Fig.~\ref{fig:naive_QDAM}). No $T$-depth optimization is thus possible: $T_{d, {\cal M}}$ has $O(N)$ and the $T$-$Cost$ becomes (more than) $O(N^{3/2})$. This indicates that the quadratic query advantage is elusive.

\begin{figure}
\centering
\includegraphics[width=0.85\textwidth]{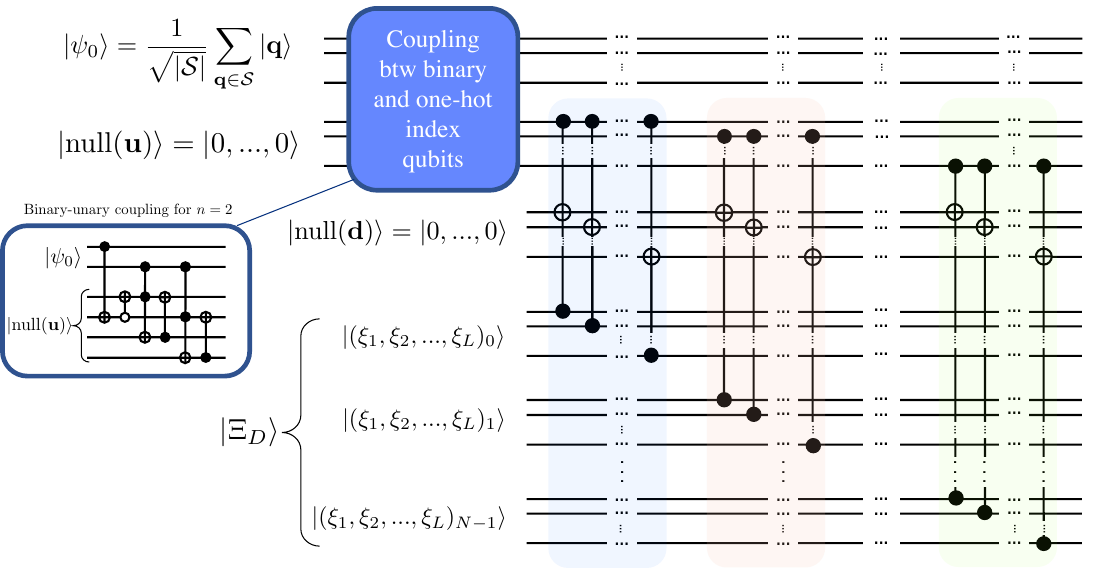}
\caption{Schematic of our $T$-depth-optimized QDAM. To reduce $T$-depth, we employ one-hot-encoded index qubits, which allows the parallelization of the Toffoli gates in data loading process. We first perform the binary- and the one-hot-encoded index coupling and then apply the data-loading process.}
\label{fig:T-optimized_QDAM} 
\end{figure}

\begin{figure}
\centering
\includegraphics[width=0.70\textwidth]{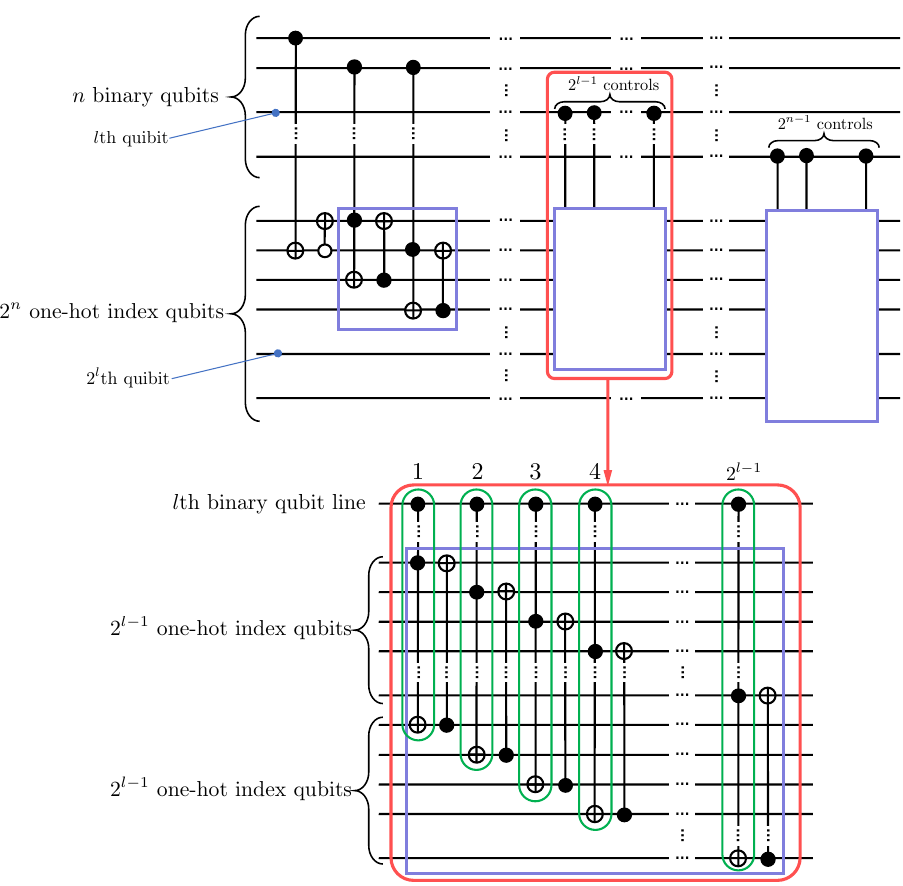}
\caption{Quantum circuit for the subprocess [{\bf M.1}] to couple between the binary- and the one-hot-encoded index qubits, implementing a generalized version of Eq. (19). This subprocess consists of $n$ steps involving each binary index qubit one by one. In each step $l=1,\cdots,N$, a total number of $2^{l-1}$ Toffoli gates using only one control qubit, i.e. the $l$th binary qubit, are applied, importantly all in parallel, as further clarified in Fig. 6. That is, the green circled parts $1,2,\cdots,2^{l-1}$ are executed in parallel, not sequentially. See Fig. 6.}
\label{fig:subprocess_M1} 
\end{figure}

\begin{figure}
\centering
\includegraphics[width=0.75\textwidth]{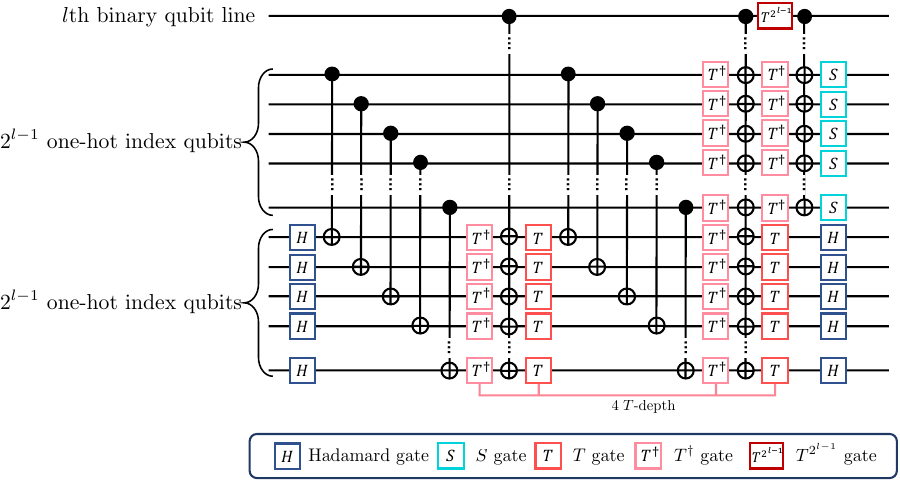}
\caption{$4$ $T$-depth optimization of the subprocess [{\bf M.1}]. A total number of $2^{l-1}$ Toffoli gates inside the red box of Fig. 5 can be implemented in parallel since they all use a single binary qubit as the control (green circled parts). Each box in Fig. \ref{fig:subprocess_M1} can thus be decomposed to $4$ $T$-depth quantum circuit. }
\label{fig:subprocess_M1-opt}
\end{figure}

\begin{figure}
\centering
\includegraphics[width=0.85\textwidth]{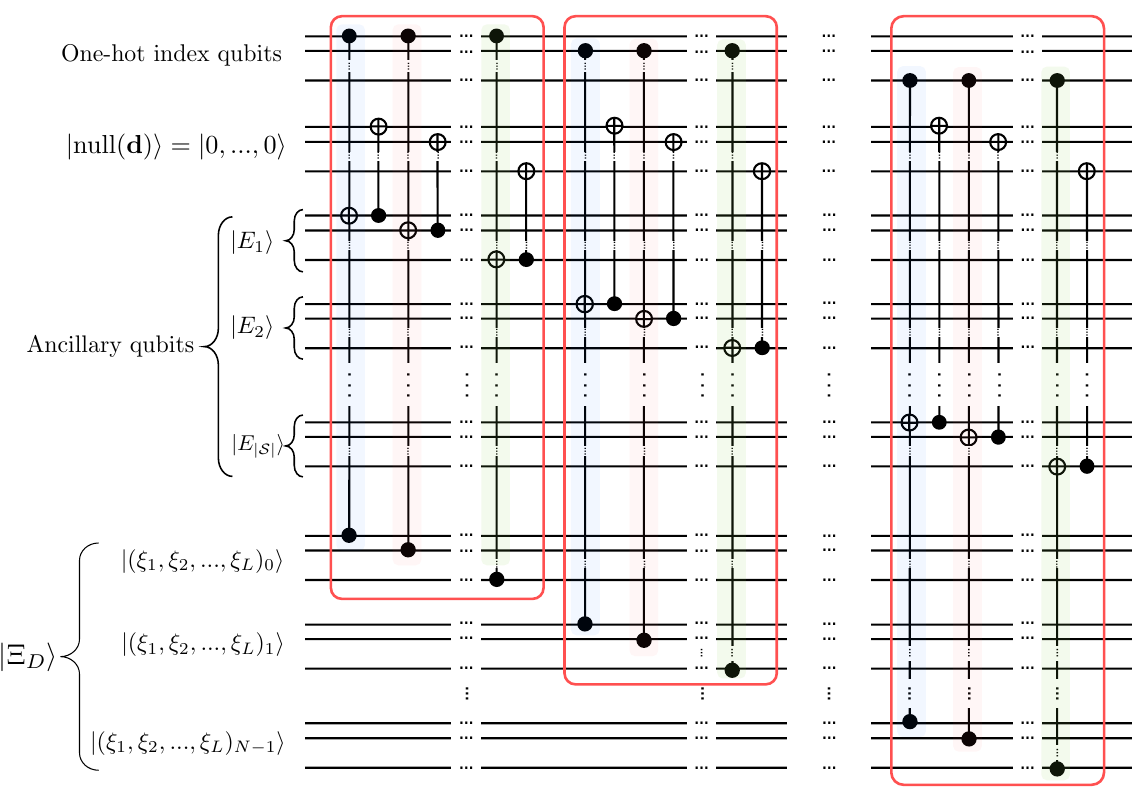}
\caption{Quantum circuit for the subprocess [{\bf M.2}] to load data. Without $T$-depth optimization, $|\mathcal S|$ number of Toffoli gates would flip each data qubit (the target qubit) sequentially. To reduce the $T$-depth, we employ ancillary qubits denoted by $\ket{E_1}, \ket{E_2}, \cdots$ that Toffoli gates flip, instead of data qubits directly, which can be executed in parallel reducing $T$-depth complexity. Then, one can subsequently load the data by sequentially applying a CNOT gate using the ancillary qubit as the control and the data qubit as the target, which do not incur the $T$-depth cost.  Therefore, the $T$-depth cost becomes $4$.}
\label{fig:subprocess_M2}
\end{figure}

We now design our $T$-depth-optimized QDAM. A crucial idea is to employ the one-hot-encoded index qubits defined in Eq.~(\ref{eq:qb_registers}), unlike the above naive approach. At first sight, using more qubits, i.e. one-hot-encoded index qubits, may seem more complex, however, it is the key to enable gate operations in parallel reducing computational complexity: The number of control qubits of the Toffoli gates in this case is $1$ because the Hamming weight of a one-hot-encoded index $\mathbf{u}=u_1 u_2 \ldots u_{2^n}$ (i.e., the number of $u_j$'s with a string of $1$ for $j \in [1,2^n]$) is $1$. This allows the Toffoli gates to work in parallel, enabling a reduction in the circuit depth. A schematic of this idea is displayed in Fig.~\ref{fig:T-optimized_QDAM}. We perform two sub-processes to realize such a QDAM: [{\bf M.1}] First, we couple the  binary- and the one-hot-encoded  index qubits. For example, with $n=2$, we have the process represented as (see the blue panel box in Fig.~\ref{fig:T-optimized_QDAM})
\begin{eqnarray}
&& \left(\gamma_0\ket{00} + \gamma_1\ket{01} + \gamma_2\ket{10} + \gamma_3\ket{11}\right)\ket{\text{null}(\mathbf{u})} \nonumber \\
&& \to \gamma_0\ket{00}\ket{0001} + \gamma_1\ket{01}\ket{0010} + \gamma_2\ket{10}\ket{0100} + \gamma_3\ket{11}\ket{1000}, 
\end{eqnarray}
with $\ket{\text{null}(\mathbf{u})} = \ket{0000}$. Such a process can be generalized to an arbitrary $n$, as shown in Fig.~\ref{fig:subprocess_M1}, where each blue-colored box contains the Toffoli gates, each conditioned on a single index qubit (red box in Fig.~\ref{fig:subprocess_M1}). A $T$-depth-optimized (local) circuit to implement the operations of the red box in Fig.~\ref{fig:subprocess_M1} can be designed by decomposing and parallelizing the Toffoli gates properly~\cite{Di2016}. 
One possible design of those circuits is given in Fig.~\ref{fig:subprocess_M1-opt}, where only 4 $T$-gates are required importantly in parallel. We can thus complete the subprocess [{\bf M.1}] with only $4\times(n-1)$ $T$-depth. [{\bf M.2}] Second, we load the data $\ket{(\xi_\alpha)_i}$ into $\ket{\mathbf{d}_i}$ and generate the correlated state, as in Eq.~(\ref{eq:Psi_1}). The circuit for this [{\bf M.2}] comprises the Toffoli and the CNOT gates, as depicted in Fig.~\ref{fig:subprocess_M2}. Here, we employ extra ancillary qubits, denoted by $\ket{E_1},  \ket{E_2}, \cdots$, to parallelize the Toffoli gates conditioned on the different control-qubits (unary hot-index qubits), as shown in Fig.~\ref{fig:subprocess_M2}. If the ancillary qubits were not employed as in Fig.~\ref{fig:T-optimized_QDAM},  the Toffoli gates that flip the data qubits cannot be parallelized and should be performed sequentially, which would require $4m\times2^n$ of the $T$-depth. 
In contrast, by using the ancillary qubits, the Tofolli gates in each red box of Fig.~\ref{fig:subprocess_M2} can be performed in parallel. Note that each sub-circuit consisting of the Toffoli and CNOT gates is equivalent to those in the red box in Fig.~\ref{fig:subprocess_M1}, and can be optimized in the same way as shown in Fig.~\ref{fig:subprocess_M1-opt}. As a result, each data loading in [{\bf M.2}] requires only four $T$-depth.

Combining the above results for [{\bf M.1}] and [{\bf M.2}], we obtain
\begin{eqnarray}
T_{d, {\cal M}} \le 4n,
\label{eq:T_dM}
\end{eqnarray}
where the inequality is introduced, as a further optimization may be possible, e.g., by using one $T$-depth Toffoli scheme with several ancillas~\cite{Selinger2013}. Such a considerable reduction in $T_{d, {\cal M}}$ is attributed to the suitable use of the one-hot-encoded index qubits and the ancillary qubits in our design.

\begin{figure}
\centering
\includegraphics[width=0.75\textwidth]{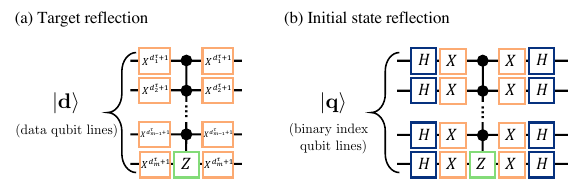}
\caption{Multiple-control-qubit $Z$ gates in the QDAM. (a) Target reflection. First, the Pauli $X$ gates are applied conditioned on the data $(\xi_\alpha)_\tau= \mathbf{d}=d_1^\tau d_2^\tau \cdots d_m^\tau$. Then, $(m-1)$-control-qubit $Z$ gate is employed. Finally, applying the Pauli $X$ gates again, one can perform the target reflection $\hat{\identity} - 2\ket{d_1^\tau d_2^\tau \cdots d_m^\tau}\bra{d_1^\tau d_2^\tau \cdots d_m^\tau}$. (b) Initial-state reflection. Generally, the initial-state reflection can be implemented in the same way as the target reflection. However, it runs on the binary index qubits, and we need $(n-1)$-control-qubit $Z$ gate. By applying the state preparation unitary and its conjugate before and after the $(n-1)$-control-qubit $Z$ gate, we can implement the initial-state reflection. The unitary to prepare $\ket{\psi_0}$ is the $n$-fold Hadamard gate $\hat{H}^{\otimes n}$ in this case.}
\label{fig:HHR} 
\end{figure}

{\em T-depth of ${\cal O}'$}---Following the formal design of the HHR, ${\cal O}'$ can be executed using an $(m-1)$-control-qubit Toffoli gate, as depicted in Fig.~\ref{fig:HHR}(a). We can mark the factor $e^{-i \pi}$ only on the state $\ket{\mathbf{d}_\tau}$ by applying either identity $\hat{\identity}$ (or equivalently, ``do nothing'' operation) or Pauli $X$ gate $\hat{\sigma}_x$ on each data qubit. 
The decision on whether to perform $\hat{\identity}$ or $\hat{\sigma}_x$ on each data qubit is determined by the bit sequences of $(\xi_\alpha)_\tau$; $\hat{\identity}$ and $\hat{\sigma}_x$ correspond to $0$ and $1$, respectively. The bits of $(\xi_\alpha)_\tau$ are encodable into ${\cal O}'$ while its index $\tau$ is still unknown. Therefore, the $T$-depth complexity of ${\cal O}'$ is determined by the decomposition of the $(m-1)$-contol-qubit Toffoli gate sandwiched between the $\hat{\identity}$ or $\hat{\sigma}_x$ gates. Formally, the $(m-1)$-control-qubit Toffoli gate can be implemented by $2(m-1)$ Toffoli gates with each 3 $T$ depth and one CNOT gate when adopting $m-2$ additional qubits (for $m > 3$). Alternatively, it can be decomposed to $O(m^2)$ Toffoli, CNOT, and single-qubit gates without using additional qubits\footnote{Refer to Section 4.3 of Ref.~\cite{Nielsen2000}.}. Consequently, it results in a polynomial $T$-depth, specifically, 
\begin{eqnarray}
T_{d,{\cal O}'} \le 3 \times 2 (m-1).
\label{eq:T_dO}
\end{eqnarray}

{\em T-depth of ${\cal D}$}---For the phase-matching $\phi=\varphi=\pi$, ${\cal D}$ can be implemented in the same way as ${\cal O}'$---by applying an $(n-1)$-control-qubit Toffoli gate on the binary index qubits (see Fig.~\ref{fig:D_data_search}). However, the single-qubit gates $\hat{\identity}$ and $\hat{\sigma}_x$ should be replaced by the Hadamard gates $\hat{H}$, as shown in Fig.~\ref{fig:HHR}(b). This process can be readily understood by
\begin{eqnarray}
&& \hat{\identity} - 2\ket{\psi_0}\bra{\psi_0} \nonumber \\
&& = \hat{H}^{\otimes n}\left( \hat{\identity} - 2\ket{00\ldots0}\bra{00\ldots0} \right)\hat{H}^{\otimes n},
\end{eqnarray}
where $\ket{\psi_0} = \hat{H}^{\otimes n} \ket{00\ldots0}$ with $\abs{\cal S}=2^n$. 
We thus obtain
\begin{eqnarray}
T_{d,{\cal D}} \le 3 \times 2 (n-1).
\label{eq:T_dD}
\end{eqnarray}


\subsection{$T$-$Cost$ and quadratic quantum speedup} 

In summary, the required $T$-depth resources for a single iteration of the quantum search are
\begin{eqnarray}
T_{d,{\cal M}} &=& O(4n), \nonumber \\
T_{d,{\cal O'}} &=& O(6m), \nonumber \\
T_{d,{\cal D}} &=& O(6n).
\end{eqnarray}
Therefore, for the natural condition $n \gg m$, we can achieve, in view of  Eq.~(\ref{eq:T_d}),
\begin{eqnarray}
T_d = O(cn),
\end{eqnarray}
where $c$ is a certain constant. Consequently, we confirm the following by recalling the query complexity ${\cal Q}$ in Eq.~(\ref{eq:q_query_adv}) and assuming $\abs{\cal S}=N$ (hence, $n=\log{N}$):
\begin{eqnarray}
\text{$T$-$Cost$} = O(\sqrt{N} \times \log{N}),
\end{eqnarray}
where the factors $\sqrt{N}$ and $\log{N}$ are from $Q$ and $T_d$, respectively. The quadratic quantum query advantage, i.e., $O(N) \to O(\sqrt{N})$, can actually be achieved with the runtime optimization as desired.

\section{Discussion}\label{sec:discussion}

We have proposed a concrete model of quantum search algorithm by designing an effective architecture of the quantum data-access process.  Without a proper design of quantum data access, the requirement for computational resources may substantially grow up above the desired quadratic reduction in the computational complexity. To address this crucial issue, we developed a useful quantum machinery called QDAM that enables the algorithm to efficiently access and utilize the data in the quantum superposed state. We introduced a runtime cost by incorporating the quantum query and the $T$-depth complexities to analyze the algorithm performance in a consistent framework. We assumed that the algorithm would run on an FTQC circuit consisting of the logical qubits in which an efficient error correction code can be embedded. Complete optimization of the QDAM verified that the quadratic quantum query reduction is indeed feasible in the form of an actual runtime speedup in quantum search. 

The crucial enabler to actualize such a runtime speed-up is our $T$-depth-optimized QDAM. While at least $O(N)$ $T$-depth would usually be required for data-loading, we developed the QDAM that has logarithmic, i.e., $O(\log{N})$ $T$-depth. For this purpose, it is necessary to employ one-hot-encoded qubits having the size as much as the register space occupied by the searching data, i.e., $O(N)$\footnote{We note that using the qubit register as much as $O(N)$ (i.e., the space size occupied by the $N$ searching data) is unavoidable in any QDAM construction recipe.}. If the $T$-depth of the QDAM would scale as $O(N)$, e.g., using the circuit in Fig.~\ref{fig:naive_QDAM}, the remarkable quadratic query advantage might be tarnished as the runtime, i.e, $T$-$Cost$, becomes $O(\sqrt{N} \times N)$. Therefore, our logarithmic $T$-depth QDAM is a key component to achieve the runtime speedup of quantum search in actual architecture. 

Another insight from our work is the selective use of the two encodings, i.e., the binary index and the one-hot index ones, within one computation. Note that we recall the binary index qubits after the QDAM  in our algorithm. This is because in general the binary encoding would be advantageous for the quantum processes, e.g., the initial-state reflection in our case, other than the data-access. We have used a similar approach in the previous work~\cite{Song2021}. Nevertheless, if we consider a local-purpose system for data search only, one may think of an algorithm that uses only the one-hot encoding, for example, by developing a useful method to prepare or process the W-type entangled state.

Our result provides insight into how an efficient quantum data-access model can interplay with other incorporating quantum algorithm achitecture for achieving the quantum advantage in the main computation to be realistically achieved. As a future research, the question on generalizing such an approach to other quantum algorithms may be investigated.

\section*{Acknowledgements}

This work was supported by the National Research Foundation of Korea (NRF-2021M3E4A1038213, NRF-2022M3E4A1077094), and the Ministry of Science, ICT and Future Planning (MSIP) by the Institute of Information and Communications Technology Planning and Evaluation grant funded by the Korean government (2019-0-00003, ``Research and Development of Core Technologies for Programming, Running, Implementing and Validating of Fault-Tolerant Quantum Computing System''). J.J.P. and J.K were supported by a KIAS Individual Grant (CG075502 and CG014604) at the Korea Institute for Advanced Study. M.S.K. acknowledges financial support from the KIAS visiting professorship, and EPSRC Grants (EP/W032643/1 and EP/Y004752/1). H.N. is supported by an NPRP Grant 13S-0205-200258 from Qatar National Research Fund.

\section*{References}

\bibliographystyle{iop}


\end{document}